\documentclass[12pt]{article}
\usepackage{graphicx}
\usepackage{epsfig}
\usepackage{amsmath}
\usepackage{amsthm}
\usepackage{amssymb}

\usepackage[round]{natbib}

\newcommand{\n}[1]{\mbox{\boldmath{$#1$}}}
\newcommand{\code}[1]{{\tt #1}}

\usepackage{amsmath}
\usepackage{color}
\usepackage{fancyvrb}

\newcommand{\pkg}[1]{{\fontseries{b}\selectfont #1}}
\let\proglang=\textsf
\newcommand{\Address}[1]{\def\@Address{#1}}
\newcommand{\Plaintitle}[1]{\def\@Plaintitle{#1}}
\newcommand{\Shorttitle}[1]{\def\@Shorttitle{#1}}
\newcommand{\Plainauthor}[1]{\def\@Plainauthor{#1}}
\newcommand{\Abstract}[1]{\def\@Abstract{#1}}
\newcommand{\Keywords}[1]{\def\@Keywords{#1}}
\newcommand{\Plainkeywords}[1]{\def\@Plainkeywords{#1}}
\newcommand{\ourpack}{\pkg{BayesVarSel }}
\DefineVerbatimEnvironment{Code}{Verbatim}{}

\author{Gonzalo Garcia-Donato\\Universidad de Castilla-La Mancha \And 
        Anabel Forte\\Universidad de Valencia}
        
\title{BayesVarSel:  Bayesian Testing, Variable Selection and model averaging in Linear Models using R}

\author{Gonzalo Garcia-Donato$^1$ and Anabel Forte$^2$\\ 
\footnotesize{$^1$ Universidad de Castilla La Mancha; $^2$ Universidad de Valencia}}


\begin{document}
\maketitle
\begin{abstract}
This paper introduces the \proglang{R} package \pkg{BayesVarSel} which implements objective Bayesian methodology for hypothesis testing and variable selection in linear models. The package computes posterior probabilities of the competing hypotheses/models and provides a suite of tools, specifically proposed in the literature, to properly summarize the results. Additionally, \ourpack\ is armed with functions to compute several types of model averaging estimations and predictions with weights given by the posterior probabilities. \ourpack\ contains exact algorithms to perform fast computations in problems of small to moderate size and heuristic sampling methods to solve large problems. The software is intended to appeal to a broad spectrum of users, so the interface has been carefully designed to be highly intuititive and is inspired by the well-known \code{lm} function. The issue of prior inputs is carefully addressed. In the default usage (fully automatic for the user) \ourpack\ implements the criteria-based priors proposed by \cite{Baetal11}, but the advanced user has the possibility of using several other popular priors in the literature. The package is available
  through the Comprehensive R Archive Network, CRAN. We illustrate the
  use of \ourpack\ with several data examples.
\end{abstract}

\section[An illustrated overview of BayesVarSel]{An illustrated overview of \pkg{BayesVarSel}}
Testing and variable selection problems are taught in almost any introductory statistical course. In this first section we assume such background to present the essence of the Bayesian approach and the basic usage of \ourpack\ with hardly any mathematical formulas. Our motivating idea in this first section is mainly to present the appeal of the Bayesian answers to a very broad spectrum of applied researchers.

The remaining sections are organized as follows. In Section 2 the problem is presented and the notation needed is introduced jointly with the basics of the Bayesian methodology. Section 3 and Section 4 explain the details concerning the obtention of posterior probabilities in hypothesis testing and variable selection problems, respectively, in \ourpack. In Section 5 several tools to describe the posterior distribution are explained and Section 6 is devoted to model averaging techniques. Finally, Section 7 concludes with directions that we plan to follow for the future of the \ourpack project. In the Appendix, formulas for the most delicate ingredient in the underlying problem in \ourpack, namely the prior distributions for parameters within each model, are collected.

\subsection{Testing}
\label{sec.ov.test}
In testing problems, several competing hypotheses, $H_i$, about a phenomenon of interest are postulated. The role of statistics is to provide summaries about the evidence in favor (or against) the hypotheses once the data, $\n y$, have been observed. There are many important statistical problems with roots in testing like model selection (or model choice) and model averaging.

The formal Bayesian response to testing problems is based on the posterior probabilities of the hypotheses that summarize, in a fully understandable way, all the available information. In \ourpack an objective point of view \citep[in the sense explained in][] {Ber06} is adopted and the reported posterior probabilities only depend on $\n y$ and the statistical model assumed. These have hence the great appeal of being fully automatic for users.

For illustrative purposes consider the nutrition problem in \citet{Lee97}, p143. There it is tested, based on the following sample of 19 weight gains (expressed in grams) of rats, 
\small
\begin{Code}
R> weight.gains<- c(134, 146, 104, 119, 124, 161, 107, 83, 113, 129, 97, 123,
+	70, 118, 101, 85, 107, 132, 94)
\end{Code}
\normalsize
whether there is a difference between the population means of the group with a high proteinic diet (the first 12) or the control group (the rest):
\small
\begin{Code}
R> diet<- as.factor(c(rep(1,12),rep(0,7)))
R> rats<- data.frame(weight.gains=weight.gains, diet=diet)
\end{Code}
\normalsize
This problem (usually known as the two-samples $t$-test) is normally written as $H_0:\mu_1=\mu_2$ versus $H_1:\mu_1\ne\mu_2$,
where it is assumed that the weight gains are normal with an unknown (but common) standard deviation. The formulas that define each of the models under the postulated hypotheses are in \proglang{R} language  
\small
\begin{Code}
R> M0<- as.formula("weight.gains~1")
R> M1<- as.formula("weight.gains~diet")
\end{Code}
\normalsize
The function to perform Bayesian tests in \ourpack  is \code{Btest} which has an intuitive and simple syntax (see Section~\ref{sec.testing} for a detailed description). In this example
\small
\begin{Code}
R> Btest(models=c(H0=M0, H1=M1), data=rats)
Bayes factors (expressed in relation to H0)
 H0.to.H0  H1.to.H0
1.0000000 0.8040127
---------
Posterior probabilities:
   H0    H1
0.554 0.446	
\end{Code}
\normalsize
We can conclude that both hypotheses are similarly supported by the data (posterior probabilities close to 0.5) so there is no evidence that the diet has any impact on the average weight. 

Another illustrative example concerns the classic dataset \code{savings} (Belsley, Kuh, and Welsch, 1980)
considered by Faraway (2002), page 29 and distributed under the package \pkg{faraway}. This dataset contains macroeconomic data on 50 different countries during 1960-1970 and the question posed is to elucidate if \code{dpi} (per-capita disposable income in U.S), \code{ddpi} (percent rate of change in per capita disposable income), population under (over) 15 (75) \code{pop15} (\code{pop75}) are all explanatory variables for \code{sr}, the aggregate personal saving divided by disposable income which is assumed to follow a normal distribution. This can be written as a testing problem about the regression coefficients associated with the variables with hypotheses
$$
H_0:\beta_{dpi}=\beta_{ddpi}=\beta_{pop15}=\beta_{pop75}=0
$$
versus the alternative, say $H_1$, that all predictors are needed. The competing models can be defined as
\small
\begin{Code}
R> fullmodel<- as.formula("sr~pop15+pop75+dpi+ddpi")
R> nullmodel<- as.formula("sr~1")
\end{Code}
\normalsize
and the testing problem can be solved with
\small
\begin{Code}
R> Btest(models=c(H0=nullmodel, H1=fullmodel), data=savings)
---------
Bayes factors (expressed in relation to H0)
H0.to.H0 H1.to.H0
  1.0000  20.9413
---------
Posterior probabilities:
    H0    H1
 0.046 0.954
\end{Code}
\normalsize
The conclusion is that there is substantial evidence favoring $H_1$, the hypothesis that all considered predictors explain the response $sr$. 

Of course, more hypotheses can be tested at the same time. For instance, a simplified version of $H_1$ that does not include pop15 is 
$$
H_2:\beta_{dpi}=\beta_{ddpi}=\beta_{pop75}=0
$$
that can be included in the analysis as
\small
\begin{Code}
R> reducedmodel<- as.formula("sr~pop75+dpi+ddpi")
R> Btest(models=c(H0=nullmodel, H1=fullmodel, H2=reducedmodel), data=savings)
Bayes factors (expressed in relation to H0)
  H0.to.H0   H1.to.H0   H2.to.H0
 1.0000000 20.9412996  0.6954594
---------
Posterior probabilities:
   H0    H1    H2
 0.044 0.925 0.031
\end{Code}
\normalsize
The result clearly evidences $H_1$ as the best explanation for the experiment among those considered.

This scenario can be extended to check which subset of  the four covariates is the most suitable one to explain $sr$. In general, the problem of selecting the best subset of covariates from a group of potential ones is better known as variable selection.

\subsection{Variable selection}\label{sec.over.vs}
Variable selection is a multiple testing problem where each hypothesis proposes a possible subset of $p$ potential explanatory variables initially considered. Notice that there are $2^p$ hypotheses, including the simplest one stating that none of the variables should be used. 

A variable selection approach to the economic example above with $p=4$ has 16 hypotheses and can be solved using the \code{Btest} function. Nevertheless, \ourpack \ has specific facilities to handle the specificities of variable selection problems. A main function for variable selection is \code{Bvs}, fully described in Section~\ref{sec.vs}. It has a simple syntax inspired by the well-known \code{lm} function. The variable selection problem in this economic example can be solved executing
\small
\begin{Code}
R> Bvs(formula="sr~pop15+pop75+dpi+ddpi", data=savings)
The 10 most probable models and their probabilities are:
   pop15 pop75 dpi ddpi        prob
1      *     *   *    * 0.297315642
2      *     *        * 0.243433493
3      *              * 0.133832367
4      *                0.090960327
5      *         *    * 0.077913429
6      *     *          0.057674755
7      *         *      0.032516780
8      *     *   *      0.031337639
9                       0.013854369
10           *        * 0.006219812
\end{Code}
\normalsize
With a first look at these results, we can see that the most probable model is the model with all covariates (probability 0.30), which is closely followed by the one without \code{dpi} with a posterior probability of 0.24.

As we will see later, a variable selection exercise generates a lot of valuable information of which the above printed information is only a very reduced summary. This can be accessed with specific methods that explore the characteristics of objects of the type created by \code{Bvs}.

\section{Basic formulae}
The problems considered in \ourpack \ concern Gaussian linear models. Consider a response variable $\n y$, size $n$, assumed to follow the linear model (the subindex $F$ refers to {\sl full} model)
\begin{equation}\label{eq.full}
M_F:\n y=\n X_0\n\alpha+\n X\n\beta+\n\varepsilon,\,\,\,\n\varepsilon\sim N_n(\n 0,\sigma^2\n I_n),
\end{equation}
where the matrices $\n X_0:n\times p_0$, $\n X:n\times p$ and the regression vector coefficients are of conformable dimensions. Suppose you want to test $H_0:\n\beta=\n 0$ versus $H_F:\n\beta\ne\n 0$, that is, to decide whether the regressors in $\n X$ actually explain the response. This problem is equivalent to the model choice (or model selection) problem with competing models $M_F$ and
\begin{equation}\label{eq.null}
M_0:\n y=\n X_0\n\alpha+\n\varepsilon,\,\,\,\n\varepsilon\sim N_n(\n 0,\sigma^2\n I_n),
\end{equation}
and we will refer to models or hypotheses indistinctly.

Posterior probabilities are based on the Bayes factors (see \citep{KassRaf95}), a measure of evidence provided by \ourpack \ when solving testing problems. The Bayes factor of $H_F$ to $H_0$ is
$$
B_{F0}=\frac{m_F(\n y)}{m_0(\n y)}
$$
where $m_F$ is the integrated likelihood or prior predictive marginal:
$$
m_0(\n y)=\int M_0(\n y\mid\n\alpha,\sigma)\,\pi_0(\n\alpha,\sigma)d\n\alpha\,d\sigma,
$$
and
$$
m_F(\n y)=\int M_F(\n y\mid\n\alpha,\n\beta,\sigma)\,\pi_F(\n\alpha,\n\beta,\sigma)d\n\beta\,d\n\alpha\,d\sigma.
$$
Above, $\pi_0$ and $\pi_F$ are the prior distributions for the parameters within each model. From an objective point of view (the one adopted here), these distributions should be fully automatic. The assignment of such priors (which we call model selection priors) is quite a delicate issue \citep[see][]{BergerPericchi01} and has inspired many important contributions in the literature.  The package \ourpack \ allows using many of the most important proposals, which are fully detailed in the Appendix. The prior implemented by default is the robust prior by \citet{Baetal11} as it can be considered optimal in many senses and is based on a foundational basis. 

Posterior probabilities can be obtained as
$$
Pr(H_F\mid\n y)=\frac{B_{F0}Pr(H_F)}{(Pr(H_0)+B_{F0}Pr(H_F))},\,\,Pr(H_0\mid\n y)=1-Pr(H_F\mid\n y),
$$
where $Pr(H_F)$ is the probability, a priori, that hypothesis $H_F$ is true.

Similar formulas can be obtained when more than two hypotheses, say $H_1,\ldots,H_N$, are tested. In this case
\begin{equation}\label{eq.post}
Pr(H_i\mid\n y)=\frac{B_{i0}(\n y)Pr(H_i)}{\sum_{j=1}^N B_{j0}(\n y)Pr(H_j)},\,\, i=1,\ldots,N
\end{equation}
which is the posterior distribution over the model space (which is the set that contains all competing models). For simplicitly, the formula in (\ref{eq.post}) has been expressed, without any loss of generality, using Bayes factors to the null model but the same results would be obtained by fixing any other model. The default definition for $Pr(H_i)$ in testing problems is to use a constant prior, which assigns the same probability to all models, that is, $Pr(H_i)=1/N$.


For instance, within the model
$$
M_3:\n y=\alpha\n 1_n+\beta_1\n x_{1}+\beta_2\n x_{2}+\n\varepsilon
$$
we cannot test the hypotheses $H_1:\beta_1=0,\beta_2\ne 0$, $H_2:\beta_1\ne 0,\beta_2=0$, $H_3:\beta_1\ne 0,\beta_2\ne 0$ since neither $M_1$ (the model defined by $H_1$) nor $M_2$ are nested in the rest. Nevertheless, it is perfectly possible to test the problem with the four hypotheses $H_1,H_2,H_3$ (as just defined) plus $H_0:\beta_1=0,\beta_2=0$, but of course $H_0$ \emph{must} be, a priori, a plausible hypothesis. In this last case $H_0$ would take the role of null model. 

Hypotheses do not have to be necessarily of the type $\n\beta=\n 0$ and, if testable (see \cite{ravdey02} for a proper definition) any linear combination of the type $\n C^t\n\beta=\n 0$ can be considered a hypothesis. For instance one can be interested in testing $\beta_1+\beta_2=0$. In \citet{BayGar07} it was formally shown that these hypotheses can be, through reparameterizations, reduced to hypotheses like $\n\beta=\n 0$. In Section~\ref{sec.testing} we show examples of how to solve these testing problems in \ourpack \ .

Variable selection is a multiple testing problem but is traditionally presented with convenient specific notation that uses a $p$ dimensional binary vector $\n\gamma=(\gamma_1,\ldots,\gamma_p)$ to identify the models. Consider the full model in \eqref{eq.full}, and suppose that $\n X_0$ contains fix covariates that are believed to be sure in the true model (by default $\n X_0=\n 1_n$ that would make the intercept present in all the models). Then each $\gamma\in\{0,1\}^p$ defines a hypothesis $H_\gamma$ stating which $\beta$'s (those with $\gamma_i=0$) corresponding to each of the columns in $\n X$ are zero. Then, the model associated with $H_\gamma$ is 
\begin{equation}\label{eq.mgamma}
M_\gamma:\n y=\n X_0\n\alpha+\n X_\gamma\n\beta_\gamma+\n\varepsilon,\,\,\,\n\varepsilon\sim N_n(\n 0,\sigma^2\n I_n),
\end{equation}
where $\n X_\gamma$ is the matrix with the columns in $\n X$ corresponding to the ones in $\n\gamma$. Notice that $\n X_\gamma$ is a $n\times p_\gamma$ matrix where $p_\gamma$ is the number of 1's in $\n\gamma$. 

Clearly, in this variable selection problem there are $2^p$ hypotheses or models and the null model is \ref{eq.null} corresponding to $\n\gamma=\n 0$. 

A particularity of variable selection is that it is affected by multiplicity issues. This is because, and specially for moderate to large $p$, the possibility of a model showing spurious evidence is high (just because many hypotheses are considered simultaneously). As concluded in \citet{ScottBerger06} multiplicity must be controlled with the prior probabilities $Pr(H_\gamma)$ and the constant prior does not control for multiplicity. Instead, these authors propose using
\begin{equation}\label{eq.sb}
Pr(H_\gamma)=\big((p+1) {p\choose p_\gamma} \big)^{-1}.
\end{equation}
The assignment above states that models of the same dimension (the dimension of $M_\gamma$ is $p_\gamma+p_0$) should have the same probability which must be inversely proportional to the number of models of that dimension. In the sequel we refer to this prior as the ScottBerger prior.

Both the ScottBerger prior and the Constant prior for $Pr(H_\gamma)$ are particular cases of the very flexible prior 
\begin{equation}\label{eq:MSprior}
Pr(M_{\gamma}\mid\theta)=\theta^{p_\gamma}(1-\theta)^{p-p_\gamma},
\end{equation}
where the hyperparameter $\theta\in(0,1)$ has the interpretation of the common probability that a given variable is included (independently of all others).

The Constant prior corresponds to $\theta=1/2$ while the ScottBerger to $\theta\sim\mbox{Unif}(0,1)$. \cite{LeySteel09} study priors for $\theta$ of the type 
\begin{equation}\label{eq:Beta1b}
\theta\sim Beta(\theta\mid 1,b).
\end{equation} 
They argue that, 
on many occasions the user has, a priori, some information regarding the number of covariates (among the $p$ initially considered) that are expected to explain the response, say $w^\star$. As they explain, this information can be translated into the analysis assigning in (\ref{eq:Beta1b}) $b=(p-w^\star)/w^\star$. The resulting prior specification has the property that the expected number of covariates is precisely $w^\star$.

Straightforward algebra shows that assuming (\ref{eq:Beta1b}) into (\ref{eq:MSprior}) is equivalent to (integrating out $\theta$)
\begin{equation}\label{eq:MSpriorb}
Pr(M_{\gamma}\mid b)\propto \Gamma(p_\gamma+1)\Gamma(p-p_\gamma+b).
\end{equation}

\section[Hypothesis testing with BayesVarSel]{Hypothesis testing with \pkg{BayesVarSel}}\label{sec.testing}
Tests are solved in \pkg{BayesVarSel} with \code{Btest} which, in its default usage, only depends on two arguments: \code{models} a named list of \code{formula}-type objects defining the models compared  and \code{data} the \code{data.frame} with the data. 

The prior probabilities assigned to hypotheses is constant, that is, $Pr(H_i)=1/N$. This default behavior can be modified specifying \code{prior.models = "User"} jointly with the argument \code{priorprobs} that must contain a named list (with names as specified in main argument \code{models}) with the prior probabilities to be used for each hypotheses.

In the last example of Section~\ref{sec.ov.test}, we can state that the simpler model is twice as likely as the other two as:
\small
\begin{Code}
R> Btest(models=c(H0=nullmodel, H1=fullmodel, H2=reducedmodel), data=savings,
prior.models="User", priorprobs=c(H0=1/2, H1=1/4, H2=1/4))
---------
Bayes factors (expressed in relation to H0)
  H0.to.H0   H1.to.H0   H2.to.H0 
 1.0000000 21.4600656  0.7017864 
---------
Posterior probabilities:
   H0    H1    H2 
0.083 0.888 0.029 
\end{Code}
\normalsize
Notice that the Bayes factor remains the same, and the change is in posterior probabilities.

\code{Btest} tries to identify the simplest model (nested in all the others) using the names of the variables. If such model does not exist, the execution of the function stops with an error message. 
Nevertheless, there are important situations where the simplest hypothesis is defined through linear restrictions (sometimes known as `testing a subspace`) making it very difficult to determine its existence just using the names. We illustrate this situation with an example.

Consider for instance the extension of the savings example in Faraway (2002), page 32 where $H_{eqp}:\beta_{pop15}=\beta_{pop75}$ is tested against the full alternative. This null hypothesis states that the effect on personal savings, \code{sr}, of both segments of populations is similar. The model under $H_{eqp}$ can be specified as
\small
\begin{Code}
R> equalpopmodel<- as.formula("sr~I(pop15+pop75)+dpi+ddpi")
\end{Code}
\normalsize
but the command
\small
\begin{Code}
R> Btest(models=c(Heqp=equalpopmodel, H1=fullmodel), data=savings)
\end{Code}
\normalsize
produces an error, although it is clear that $H_{eqp}$ is nested in $H_1$. To overcome this error, the user must ask the \code{Btest} to relax the names-based check defining as \code{TRUE} the  argument \code{relax.nest}. In our example
\small
\begin{Code}
R> Btest(models=c(Heqp=equalpopmodel, H1=fullmodel), data=savings, 
+  relax.nest=TRUE)
Bayes factors (expressed in relation to Heqp)
Heqp.to.Heqp   H1.to.Heqp 
  1.0000000    0.3336251 
---------
Posterior probabilities:
Heqp   H1 
0.75 0.25 
\end{Code}
\normalsize
Now \code{Btest} identifies the simpler model as the one with the largest sum of squared errors and trusts the user on the existence of a simpler model (yet the code produces an error if it detects that the model with the largest sum of squared errors is not of a smaller dimension than all the others in which case it is clear that a null model does not exist). 

\section[Variable selection with BayesVarSel]{Variable selection with \pkg{BayesVarSel}}\label{sec.vs}
The number of entertained hypotheses in a variable selection problem, $2^p$, can range from a few to an extremely large number. This makes necessary to program specific tools to solve the multiple testing problem in variable selection problems. \ourpack \ provides three different functions for variable selection
\begin{itemize}
\item \code{Bvs} performs exhaustive enumeration of hypotheses and hence the size of problems must be small or moderate (say $p\le 25$),
\item \code{PBvs} is a parallelized version of \code{Bvs} making it possible to solve moderate problems (roughly the same size as above) in less time with the help of several cpu's.
\item \code{GibbsBvs} simulations from the the posterior distribution over the model space using a Gibbs sampling scheme (intended to be used for large problems, with $p>25$).
\end{itemize}

Except for a few arguments that are specific to the algorithm implemented (eg. the number of cores in \code{PBvs} or the number of simulations in \code{GibbsBvs}) the usage of the three functions is very similar. We describe the  common use in the first of the following sub-sections and the function-specific arguments in the second.  

These three functions return objects of class \code{Bvs} which are a \code{list} with relevant information about the posterior distribution. For these objects \ourpack \ provides a number of functions, based on the tradition of model selection methods, to summarize the corresponding posterior distribution (eg. what is the hypothesis most probable a posteriori) and for its posterior usage (eg. to obtain predictions or model averaged estimates). These capabilities are described in Section~\ref{sec.summaries} and Section~\ref{sec.ma}.

For illustrative purposes we use the following datasets:

\paragraph{{\tt USCrime} data.} The US Crime data set  was first studied by \cite{E73} and is available from R-package {\tt MASS} \citep{MASS}. This data set has a total of $n=47$  observations (corresponding to states in the US) of $p=15$ potential covariates aimed at explaining the rate of crimes in a particular category per head of population (labelled {\tt y} in the data).

\paragraph{{\tt SDM} data.} This dataset has a total of $p=67$ potential drivers for the annual GDP growth per capita between 1960 and 1996 for $n = 88$ countries (response variable labelled {\tt y} in the data). This data set was initially considered by \cite{SDM} and revisited by \cite{LeySteel07}.

\subsection{Common arguments}\label{sec.vs.common}
The customary arguments in \code{Bvs}, \code{PBvs} and \code{GibbsBvs} are \code{data} (a \code{data.frame} with the data) and \code{formula}, with a definition of the most complex model considered (the full model in \eqref{eq.full}). The default execution setting  corresponds to a problem where the null model (\ref{eq.null}) contains just the intercept (ie $\n X_0=\n 1_n$) and prior probabilities for models are defined as in \eqref{eq.sb}. 

A different simpler model can be specified with the optional argument \code{fixed.cov}, a character vector with the names of the covariates included in the null model. Notice that, by definition, the variables in the null model are part of any of the entertained models including of course the full model. A case sensitive convention here is to use the word ``Intercept'' to stand for the name of the intercept so the default corresponds to \code{fixed.cov=c("Intercept")}. A null model that just contains the error (that is, $\n X_0$ is the null matrix) 
is specified as \code{fixed.cov=NULL}.

Suppose for example that in the UScrime dataset and apart from the constant, theory suggests that the covariate \code{Ed} must be used to explain the dependent variable. To consider these conditions we execute the command
\small
\begin{Code}
R> crime.Edfix<- Bvs(formula="y~.", data=UScrime, fixed.cov=c("Intercept", "Ed"))
Info. . . .
Most complex model has 16 covariates
From those 2 are fixed and we should select from the remaining 14 
M, So, Po1, Po2, LF, M.F, Pop, NW, U1, U2, GDP, Ineq, Prob, Time
The problem has a total of 16384 competing models
Of these, the  10 most probable (a posteriori) are kept
Working on the problem...please wait.
\end{Code}
\normalsize
\normalsize
During the execution (which takes about 0.22 seconds in a standard laptop) the function informs which variables take part of the selection process. The number of these defines $p$ which in this problem is $p=14$ (and the model space has $2^{14}$ models). In what follows, and unless otherwise stated we do not reproduce this informative output to save space.

The assignment of priors probabilities, $Pr(H_i)$, is regulated with the argument \code{prior.models} which by default takes the value ``ScottBerger'' that corresponds to the proposal in \eqref{eq.sb}. Other options for this argument are ``Constant'', which stands for $Pr(H_i)=1/2^p$, and the more flexible value, ``User'', under which the user must specify the prior probabilities with the extra argument \code{priorprobs}.

The argument \code{priorprobs} is a $p+1$ numeric vector, which in its $i$-th position defines the probability of a model of dimension $p_0+i-1$ (these probabilities can be specified except for the normalizing constant). 

Suppose that in the UScrime with null model just the intercept, we want to specify the prior in eq~\ref{eq:MSprior} with $\theta=1/4$, this can be done as (notice that here $p=15$)
\small
\begin{Code}
R> theta<- 1/4; pgamma<- 0:15
R> crime.thQ<- Bvs(formula="y~.", data=UScrime, prior.models="User", 
+  priorprobs=theta^pgamma*(1-theta)^(15-pgamma))
\end{Code}
\normalsize

In variable selection problems it is quite standard to have the situation where the number of covariates is large (say larger than 30) preventing the exhaustive enumeration of all the competing models. The SDM dataset is an example of this situation with $p=67$. 
In these contexts, the posterior distribution can be explored using the function \code{GibbsBvs}. To illustrate the elicitation of prior probabilities as well, suppose that the number of expected covariates to be present in the true model is $w^\star=10$. This situation is considered in \cite{LeySteel09} and can be implemented as  (see eq.\ref{eq:MSpriorb}) 
\small
\begin{Code}
R> set.seed(1234)
R> wstar<- 7; b<- (67-wstar)/wstar; pgamma<- 0:67
R> growth.wstar7<- GibbsBvs(formula="y~.", data=SDM, prior.models="User", 
+  priorprobs=gamma(pgamma+1)*gamma(67-pgamma+b))				
\end{Code}
\normalsize
The above code took 18 seconds to run.

One last common argument to \code{Bvs}, \code{PBvs} and \code{GibbsBvs} is \code{time.test}. If it is set to \code{TRUE} and the problem is of moderate size ($p\ge 18$ in \code{Bvs}, \code{PBvs} and $p\ge 21$ in \code{GibbsBvs}), an estimation of computational time is calculated and the user is asked about the possibility of not executing the command.

\subsection{Specific arguments}
\paragraph{In \code{Bvs}} The algorithm implemented in \code{Bvs} is exact in the sense that the information collected about the posterior distribution takes into account {\em all} competing models as these are all computed. Nevertheless, to save computational time and memory it is quite appropriate to keep only a moderate number of the best (most probable a posteriori) models. This number can be specified with the argument \code{n.keep} which must be an integer number between 1 (only the most probable model is kept) and $2^p$ (a full ordering of models is kept). The default value for \code{n.keep} is 10. 

The argument \code{n.keep} is not of great importance to analyze the posterior distribution over the model space. Nevertheless, it has a more relevant effect if model averaging estimates or predictions are to be obtained (see Section~\ref{sec.ma}) since, as \ourpack \ is designed, only the \code{n.keep} retained models are used for these tasks.

\paragraph{In \code{PBvs}} This function conveniently distributes several \code{Bvs} among the number of available cores specified in the argument \code{n.nodes}. Another argument in \code{PBvs} is \code{n.keep} explained above.

\paragraph{In \code{GibbsBvs}} The algorithm in \code{GibbsBvs} samples models from the posterior over the model space
and this is done using a simple (yet very efficient) Gibbs sampling scheme introduced in \cite{GeorgeMcCulloch97}, later studied in \cite{Ga-DoMa-Be13} in the context of large model spaces. The type of default arguments that can be specified in \code{GibbsBvs} are the typical in any Monte Carlo Markov Chain scheme (as usual the default values are given in the assignment)
\begin{itemize}
  \item \code{init.model="Full"} The model at which the simulation process starts. Options include "Null" (the model only with the covariates specified in \code{fixed.cov}), "Full" (the model defined by \code{formula}), "Random" (a randomly selected model) and a vector with $p$ zeros and ones defining a model.
  \item \code{n.burnin=50} Length of burn in, i.e. the number of iterations to discard at the start of the simulation process.
  \item \code{n.iter=10000} The total number of iterations performed after the burn in process. 
  \item \code{n.thin=1} Thinning rate that must be a positive integer.  Set 'n.thin' > 1 to save memory and computation time if 'n.iter' is large.
  \item \code{seed=runif(1, 0, 16091956)} A seed to initialize the random number generator. (Doesn't it seem like the upper bound is a date?)
\end{itemize}

Notice that the number of total iterations is \code{n.burnin+n.iter} but the number of models that are used to collect information from the posterior is, approximately, \code{n.iter/n.thin}.

\section{Summaries of the posterior distribution}\label{sec.summaries}
In this section we describe the tools implemented in \ourpack \ conceived to summarize, in the tradition of the relevant model selection literature, the posterior distribution over the model space. In \proglang{R} this corresponds to describing methods to explore the content of objects of class \code{Bvs}. 

Printing a \code{Bvs} object created with \code{Bvs} or \code{PBvs} shows the best 10 models with their associated probability (see examples in Section~\ref{sec.over.vs}). If the object was built with  \code{GibbsBvs} then what is printed is the most probable model among the sampled ones. For instance, if we print the object \code{growth.wstar7} in Section~\ref{sec.vs.common} we obtain the following message:
\small
\begin{Code}
R> growth.wstar7
Among the visited models, the model with the largest probability contains: 
[1] "DENS65C"  "EAST"     "GDPCH60L" "IPRICE1"  "P60"      "TROPICAR"
\end{Code}
\normalsize
The model that contains the variables above plus the intercept is a point estimate of the Highest Posterior Probability model (HPM).

The rest of the summaries are very similar independently of the routine used to create it, but recall that if the object was obtained with either \code{Bvs} and \code{PBvs} (likely because $p$ is small or moderate) the given measures here explained are {\em exact}. If instead \code{GibbsBvs} was used, the reported measures are approximations of the exact ones (that likely cannot be computed due to the huge size of the model space). In \ourpack \ these approximations are based on the frequency of visits as an estimator of the real $Pr(M_\gamma\mid\n y)$ since, as studied in \cite{Ga-DoMa-Be13}, these provide quite accurate results.

The HPM is returned when an object of class \code{Bvs} is summarized (via \code{summary}) jointly with the inclusion probabilities for each competing variable, $Pr(x_i\mid\n y)$. These are the sum of the posterior probabilities of models containing that covariate and provide evidence about the individual importance of each explanatory variable. The model defined by those variables with an inclusion probability greater than 0.5 is called a Median Probability Model (MPM), which is also included in the summary. \cite{BarBer04} show that, under general conditions, if a single model has to be utilized with predictive purposes, the MPM is optimal. 

For instance, if we summarize the object \code{crime.Edfix}\footnote{Notice that variable \code{Ed} is not on the list as it was assumed to be fixed.} of Section~\ref{sec.vs.common}, we obtain
\small
\begin{Code} 
R> summary(crime.Edfix)

Inclusion Probabilities:
     Incl.prob. HPM MPM
M        0.6806       *
So       0.2386        
Po1      0.8489   *   *
Po2      0.3663        
LF       0.2209        
M.F      0.3184        
Pop      0.2652        
NW       0.2268        
U1       0.2935        
U2       0.4765        
GDP      0.3204        
Ineq     0.9924   *   *
Prob     0.6174       *
Time     0.2434        
\end{Code}
\normalsize
 We clearly see that, marginally, \code{Ineq} is very relevant followed by \code{Po1}. Less influential but of certain importance are \code{M} and \code{Prob}.

\paragraph{Graphical summaries and jointness} The main graphical support in \ourpack \ is contained in the function \code{plotBvs} which depends on \code{x} (an object of class \code{Bvs}) and the argument \code{option} which specified the type of plot to be produced:
\begin{itemize}
\item \code{option="joint"} A matrix plot with the joint inclusion probabilities, $Pr(x_i,x_j\mid\n y)$ (marginal inclusion probabilities in the diagonal).

\item \code{option="conditional"} A matrix plot with the conditional inclusion probabilities $Pr(x_i\mid x_j,\n y)$ (ones in the diagonal).

\item \code{option="not"} A matrix plot with the conditional inclusion probabilities $Pr(x_i\mid {\mbox{Not } x_j}, \n y)$ (zeroes in the diagonal).

\item \code{option="dimension"} A bar plot representation of the posterior distribution of the dimension of the true model (number of variables, ranging from $p_0$ to $p_0+p$).

\end{itemize} 

\begin{figure}[t!]
\centering
\begin{tabular}{cc}
\includegraphics[width=0.45\textwidth]{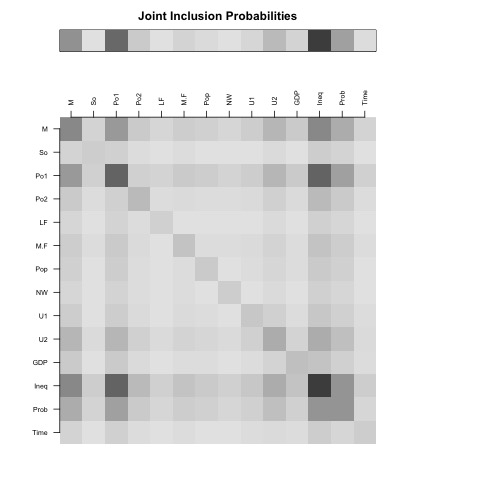}&
\includegraphics[width=0.45\textwidth]{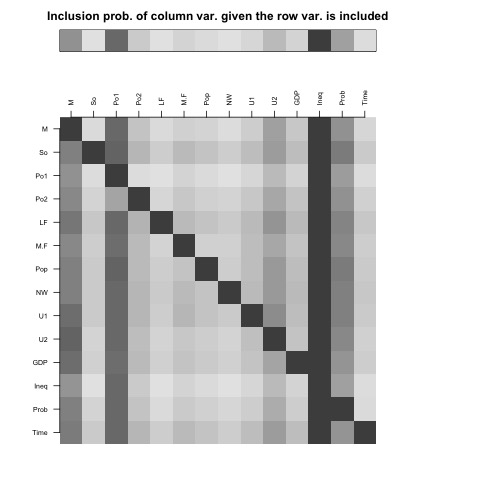}\\
\includegraphics[width=0.45\textwidth]{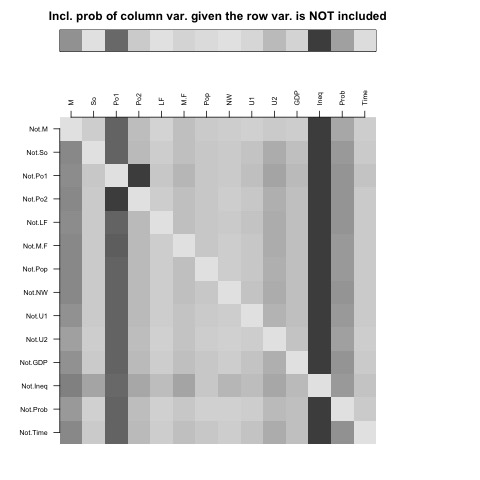}&
\includegraphics[width=0.45\textwidth]{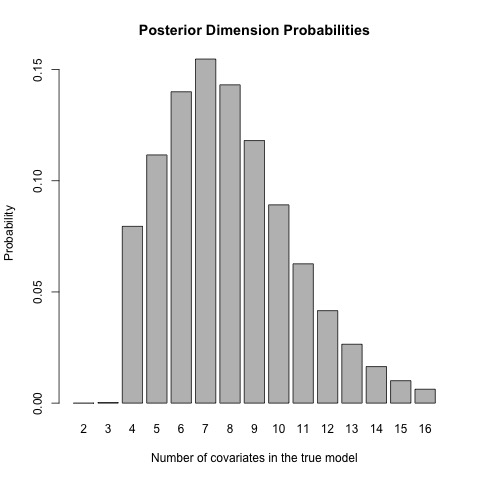}
\end{tabular}
\caption{\small Plots corresponding to the four possible values of the argument \code{option} in \code{plotBvs} over the object \code{crime.Edfix} of Section~\ref{sec.vs.common}. From left to right: \code{joint}, \code{conditional}, \code{not} and \code{dimension}}.
\label{fig:plotBvs}
\end{figure}

The first three options above are basic measures describing aspects of the joint effect of two given variables, $x_i,x_j$ and can be understood as natural extensions of the marginal inclusion probabilities. In Figure~\ref{fig:plotBvs}, we have reproduced the first three plots (from left to right) obtained with the following lines of code:
\small
\begin{Code}
R> mj<- plotBvs(crime.Edfix, option="joint")
R> mc<- plotBvs(crime.Edfix, option="conditional")
R> mn<- plotBvs(crime.Edfix, option="not")
\end{Code}
\normalsize
Apart from the plot, these functions return the matrix represented for futher study. For the conditionals probabilities (\code{conditional} and \code{not}) the variables in the row are the conditioning variables (eg. in \code{mc} above, the position $(i,j)$ is the inclusion probability of variable in $j$-th column  conditional on the variable in $i$-th row).

Within these matrices, the most interesting results correspond to variations from the marginal inclusion probabilities (represented in the top of the plots as a separate row for reference). Our experience suggests that the most valuable of these is \code{option="not"} as it can reveal key details about the relations between variables in relation with the response. For instance, take that plot in Figure~\ref{fig:plotBvs} (plot on the left of the second row) and observe that, while variable \code{Po2} barely has any effect on savings, it becomes relevant if \code{Po1} is removed. This is the probability $Pr(Po2\mid\mbox{Not}\, Po1,\n y)$ with value
\small
\begin{Code}
R> mn["Not.Po1","Po2"]
[1] 0.9996444
\end{Code}
\normalsize
which, as we observed in the graph, is substantially large compared with $Pr(Po2\mid\n y)=0.3558$ (a number printed in the summary above).

Similarly, we observe that \code{Po1} is of even more importance if \code{Po2} is not considered as a possible explanatory variable. All this implies that, in relation with the dependent variable, both variables contain similar information and one can act as proxy for the other.

We can further investigate this idea of a relationship between two covariates with respect to the response using the {\em jointness} measures proposed by \cite{LeySteel07}.  These are available using function \code{Jointness} that depends on two arguments: \code{x}, an object of class \code{Bvs} and \code{covariates} a character vector indicating which pair of covariates we are interested in. By default \code{covariates="All"} printing the matrices with the jointness measurement for every pair of  covariates.  In particular, three jointness measures relative to two covariates are reported by this function: i) the joint inclusion probability, ii) the ratio between the joint inclusion probability and the probability of including at least one of them and finally, iii) the ratio between the joint inclusion probability and the probability of including one of them alone.

For instance:
\small
\begin{Code}
R> Jointness(crime.Edfix, covariates=c("Po1","Po2"))
---------
The joint inclusion probability for Po1 and Po2 is:  0.22 
---------
The ratio between the probability of including both covariates and the probability 
of including at least one of then is: 0.22
---------
The probability of including both covariates together is 0.27 times the probability 
of including one of them alone 
\end{Code}
\normalsize
With these results we must conclude that it is unlikely that both variables, \code{Po1} and \code{Po2}, are to be included together in the true model.

Finally, within \code{plotBvs}, the assignment \code{option="dimension"} produces a plot that speaks about the complexity of the true model in terms of the number of covariates that it contains. The last plot in Figure~\ref{fig:plotBvs} is the output of
\small
\begin{Code}
R> plotBvs(crime.Edfix, option="dimension")
\end{Code}
\normalsize
From this plot we conclude that the number of covariates is about 7 but with a high variability. The exact values of this posterior distribution are in the component \code{postprobdim} of the \code{Bvs} object. 

\section{Model averaged estimations and predictions}\label{sec.ma}
In a variable selection problem it is explicitly recognized that there is uncertainty regarding which variables make up the true model. Obviously, this uncertainty should be propagated in the inferential process (as opposed to  inferences using just one model) to produce more reliable and accurate estimations and predictions. These type of procedures are normally called model averaging and are performed once the model selection exercise is performed (that is, the posterior probabilities have been already obtained). In \ourpack \ these inferences can be obtained acting over objects of class \code{Bvs}. 

Suppose that $\Lambda$ is a quantity of interest and that under model $M_\gamma$ it has a posterior distribution $\pi^N(\Lambda\mid\n y,M_\gamma)$ with respect to certain non-informative prior $\pi^N_\gamma$. Then, we can average over all entertained models using the posterior probabilities in (\ref{eq.post}) as weights to obtain
\begin{equation}\label{eq.ma}
f(\Lambda\mid\n y)=\sum_\gamma\, \pi^N(\Lambda\mid\n y,M_\gamma)\, Pr(M_\gamma\mid\n y).
\end{equation}
In \ourpack \ for $\pi^N_\gamma$ we use the reference prior developed in \cite{BerBer92} and further studied in \cite{BerBerDon09}. This is an objective prior with very good theoretical properties and the formulas for the posterior distribution with a fixed model are known \citep{BerSmi94}. These priors are different from the {\em model selection} priors used to compute the Bayes factors (see Section 2), but as shown in \cite{Consonni2016}, the posterior distributions approximately coincide and then $f(\Lambda\mid\n y)$ basically can be interpreted as the posterior distribution of $\Lambda$.

There are two different quantities $\Lambda$ that are of main interest in variable selection problems. First is a regression parameter $\beta_i$ and second is a future observation $y^\star$ associated with known values of the covariates $\n x^\star$. In what follows we refer to each of these problems as (model averaged) estimation and prediction to which we devote the next subsections.

\subsection{Estimation}
Inclusion probabilities $Pr(x_i\mid\n y)$ can be roughly interpreted as the probability that $\beta_i$ is different from zero. Nevertheless, it does not say anything about the magnitude of the coefficient $\beta_i$ nor anything about its sign. 

Such type of information can be obtained from the distribution in (\ref{eq.ma}) which in the case of $\Lambda\equiv(\n\alpha,\n\beta)$ is
\begin{equation}\label{eq.ma.est}
f(\n\alpha,\n\beta\mid\n y)=\sum_\gamma\, St_{p_\gamma+p_0}((\n\alpha,\n\beta_\gamma)\mid (\hat{\n\alpha},\hat{\n\beta}_\gamma),(\n Z_\gamma^t\n Z_\gamma)^{-1}\frac{SSE_\gamma}{n-p_\gamma-p_0},n-p_\gamma-p_0)\, Pr(M_\gamma\mid\n y),
\end{equation}
where $\hat{\n\alpha},\hat{\n\beta}_\gamma$ is the maximum likelihood estimator under $M_\gamma$ (see eq.\ref{eq.mgamma}), $\n Z_\gamma=(\n X_0,\n X_\gamma)$ and $SSE_\gamma$ is the sum of squared errors in $M_\gamma$. Above $St$ makes reference to the multivariate student distribution:
$$
St_k(\n x\mid\n\mu,\n\Sigma,df)\propto \big(1+\frac{1}{df}(\n x-\n\mu)^t\n\Sigma^{-1}(\n x-\n\mu)\big)^{-(df+k)/2}.
$$

In \ourpack \ the whole model averaged distribution in (\ref{eq.ma.est}) is provided in the form of a random sample through the function\code{BMAcoeff} which depends on two arguments: \code{x} which is a \code{Bvs} object and \code{n.sim} the number of observations to be simulated (taking the default value of 10000). 
The returned object is an object of class \code{bma.coeffs} which is a column-named \code{matrix} with \code{n.sim} rows (one per each simulation) and $p+p_0$ columns (one per each regression parameter). The way that \code{BMAcoeff} works depends on whether the object was created with \code{Bvs} (or \code{PBvs}) or with \code{GibbsBvs}. This is further explained below.

\paragraph{If the \code{Bvs} object was generated with \code{Bvs} or \code{PBvs}} In this case the models over which the average is performed are the \code{n.keep} (previously specified) best models. Hence, if \code{n.keep} equals $2^p$ then all competing models are used while if \code{n.keep}<$2^p$ only a proportion of them are used and posterior probabilities are re-normalized to sum one.

On many occasions where estimations are required, the default value of \code{n.keep} (which we recall is 10) is small and should be increased. Ideally $2^p$ should be used but, as noticed by \cite{Rafteryetal97} this is normally unfeasible and commonly it is sufficient to average over a reduced set of good models that accumulates a reasonable posterior mass. This set is what \cite{Rafteryetal97} call the ``Occam's window''. The function \code{BMAcoeff} informs about the total probability accumulated in the models that are used.

For illustrative purposes let us retake the UScrime dataset and, in particular, the example in Section~\ref{sec.over.vs} in which, apart from the constant, the variable Ed was assumed as fixed. The total number of models is $2^{14}=16384$ and we execute again \code{Bvs} but now with \code{n.keep}=2000
\small
\begin{Code}
R> crime.Edfix<- Bvs(formula="y~.", data=UScrime, 
+ fixed.cov=c("Intercept", "Ed"), n.keep=2000)
\end{Code}
\normalsize
(taking 1.9 seconds). The object \code{crime.Edfix} contains identical information as  the one previously created with the same name, except for the models retained, which in this case are the best 2000. These models accumulate a probability of 0.90, which seems quite reasonable to derive the estimates. We do so executing the second command of the following script (the seed is fixed for the sake of reproducibility).
\small
\begin{Code}
R> set.seed(1234)
R> bma.crime.Edfix<- BMAcoeff(crime.Edfix)
Simulations obtained using the best 2000 models
that accumulate 0.9 of the total posterior probability
\end{Code}
\normalsize
The distribution in (\ref{eq.ma.est}) and hence the simulations obtained can be highly multimodal and providing default summaries of it (like the mean or standard deviation) is potentially misleading. For a first exploration of the model averaged distribution, \ourpack \ comes armed with a plotting function, \code{histBMA}, that produces a histogram-like representation borrowing ideas from \cite{ScottBerger06} and placing a bar at zero with height proportional to the number of zeros obtained in the simulation. 

The function \code{histBMA} depends on several arguments:
\begin{itemize}
\item \code{x} An object of class \code{bma.coeffs}.

\item \code{covariate} A text specifying the name of an explanatory variable whose accompanying coefficient is to be represented. This must be the name of one of the columns in \code{x}.

\item \code{n.breaks} The number of equal lentgh bars for the histogram. Default value is 100.

\item \code{text} If set to \code{TRUE} (default value) the frequency of zeroes is added at the top of the bar at zero.
  
\item \code{gray.0} A numeric value between 0 and 1 that specifies the darkness, in a gray scale (0 is white and 1 is black) of the bar at zero. Default value is 0.6.

\item \code{gray.no0} A numeric value between 0 and 1 that specifies the darkness, in a gray scale (0 is white and 1 is black) of the bars different from zero. Default value is 0.8.
\end{itemize}
%

\begin{figure}[t!]
\centering
\begin{tabular}{ccc}
\includegraphics[width=0.3\textwidth]{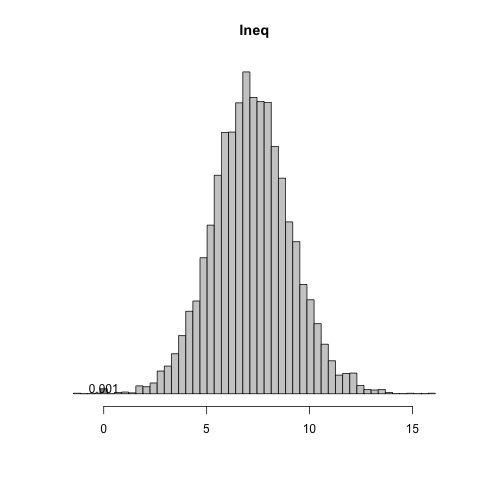}&
\includegraphics[width=0.3\textwidth]{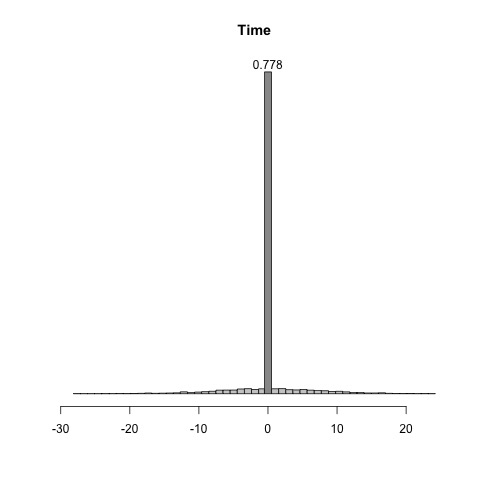}&
\includegraphics[width=0.3\textwidth]{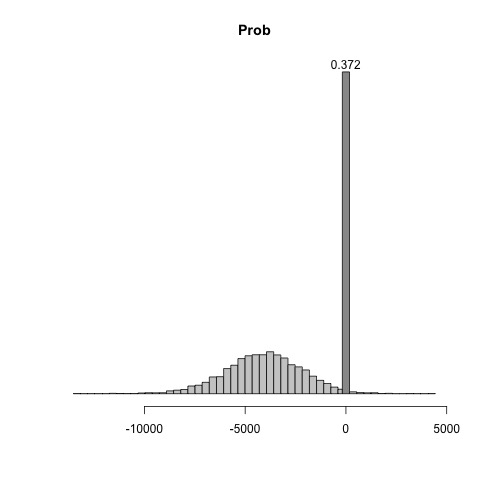}
\end{tabular}
\caption{\small Representation provided by the function \code{histBMA} of the Model averaged posterior distributions of $\beta_{Ineq}$, $\beta_{Time}$and $\beta_{Prob}$ for the UScrime dataset with constant and Ed considered as fixed in the variable selection exercise.}
\label{fig.plotbma}
\end{figure}

For illustrative purposes let us examine the distributions of $\beta_{Ineq}$ (inclusion probability 0.99); $\beta_{Time}$ (0.24) and $\beta_{Prob}$ (0.62) using \code{histBMA}
\small
\begin{Code}
R> histBMA(bma.crime.Edfix, covariate = "Ineq", n.breaks=50)
R> histBMA(bma.crime.Edfix, covariate = "Time", n.breaks=50)
R> histBMA(bma.crime.Edfix, covariate = "Prob", n.breaks=50)
\end{Code}
\normalsize
The plots obtained are reproduced in Figure~\ref{fig.plotbma} where we can see that \code{Ineq} has a positive effect. This distribution is unimodal so there is no drawback to summarizing the effect of the response \code{Ineq} over savings using, for example, the mean and quantiles, that is:
\small
\begin{Code}
R> quantile(bma.crime.Edfix[,"Ineq"], probs=c(0.05, 0.5, 0.95))
       5%       50%       95% 
 4.075685  7.150184 10.326606 
\end{Code}
\normalsize
This implies an estimated effect of 7.2 with a 90\% credible interval [4.1,10.3]. The situation of \code{Time} is clear and its estimated effect is basically null (in agreement with a low inclusion probability). 

Much more problematic is reporting estimates of the effect of \code{Prob} with a highly polarized estimated effect being either very negative (around -4100) or zero (again in agreement with its inconclusive inclusion probabilty of 0.62). Notice that, in this case, the mean (approximately -2500) should not be used as a sensible estimation of the parameter $\beta_{Prob}$.

\paragraph{If the \code{Bvs} object was generated with \code{GibbsBvs}} In this case, the average in 
(\ref{eq.ma.est}) is performed over the \code{n.iter} (an argument previously defined) models sampled in the MCMC scheme. 
Theoretically this corresponds to sampling over the whole distribution (all models are considered) and leads to the approximate  method pointed out in \cite{Rafteryetal97}. All previous considerations regarding the difficult nature of the underlying distribution apply here.

Let us consider again the SDM dataset in which analysis we created the object \code{growth.wstar7} in Section~\ref{sec.vs}. Suppose we are interested in the effect of the variable \code{P60} on the response GDP. The summary method informs that this variable has an inclusion probability of 0.77.
\small
\begin{Code}
R> set.seed(1234)
R> bma.growth.wstar7<- BMAcoeff(growth.wstar7)
Simulations obtained using the  10000  sampled models.
Their frequencies are taken as the true posterior probabilities
R> histBMA(bma.growth.wstar7, covariate = "P60",n.breaks=50)
\end{Code}
\normalsize
The distribution is bimodal (graph not shown here to save space) with modes at zero and 2.8 approximately. Again, it is difficult to provide simple summaries to describe the model averaged behaviour of P60. Nevertheless, it is always possible to answer relevant questions such as: what is the probability that the effect of \code{P60} over savings is greater than one?
\small
\begin{Code}
R> mean(bma.growth.wstar7[,"P60"]>1)
[1] 0.7511
\end{Code}
\normalsize

\subsection{Prediction} Suppose we want to predict a new observation $y^\star$ with associated values of covariates $(\n x^\star)^t\in{\cal R}^{p_0+p}$ (in what follows the product of two vectors corresponds to the usual scalar product). In this case, the distribution (\ref{eq.ma}) adopts the form
\begin{equation}\label{eq.ma.pred}
f(y^\star\mid\n y,\n x^\star)=\sum_\gamma\, St(y^\star\mid \n x^\star_\gamma\,\, (\hat{\n\alpha}, \hat{\n\beta}_\gamma), \frac{SSE_\gamma}{h_\gamma}, n-p_\gamma-p_0)\, Pr(M_\gamma\mid\n y),
\end{equation}
where
$$
h_\gamma=1-\n x^\star_\gamma\big((\n x^\star_\gamma)^t\n x^\star_\gamma+\n Z_\gamma^t\n Z_\gamma\big)^{-1}(\n x^\star_\gamma)^t.
$$
As with estimations, \ourpack \ has implemented the function \code{predictBvs} designed to simulate a desired number of observations from (\ref{eq.ma.pred}). A main difference with model averaged estimations is that, normally, the above predictive distribution is unimodal.


The function \code{predictBvs} depends on \code{x}, an object of class \code{Bvs}, \code{newdata}, a \code{data.frame} with the values of the covariates (the intercept, if needed, is automatically added) and \code{n.sim} the number of observations to be simulated. The considerations described in the previous section for the \code{Bvs} object about the type of function originally used apply here.

The function \code{predictBvs} returns a matrix with \code{n.sim} rows (one per each simulated observation) and with the number of columns the number of cases (rows) in the \code{newdata}. 

\begin{figure}[t!]
\centering
\begin{tabular}{c}
\includegraphics[width=0.45\textwidth]{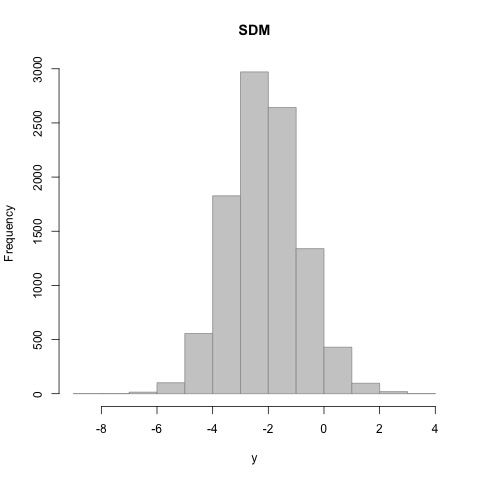}
\end{tabular}
\caption{\small For SDM data and related \code{Bvs} object \code{growth.wstar7}, model averaged prediction of the ``mean'' case (predicting the output associated with the mean of observed covariates)}
\label{fig.plotpred}
\end{figure}

For illustrative purposes, consider the \code{Bvs} object named \code{growth.wstar7} for the analysis of the SDM dataset. Simulations from the predictive distribution (\ref{eq.ma.pred}) associated with values of the covariates fixed at their means can be obtained with the following code. Here, a histogram is produced (see Figure~\ref{fig.plotpred}) as a graphical approximation of the underlying distribution.
\small
\begin{Code}
R> set.seed(1234)
R> pred.growth.wstar7<- predictBvs(x=growth.wstar7,
+  newdata=data.frame(t(colMeans(SDM))))
R> hist(pred.growth.wstar7[,1], main="SDM", 
+  border=gray(0.6), col=gray(0.8), xlab="y")
\end{Code}
\normalsize

\section{Future work} The first version of \ourpack \ was released on December 2012 with the main idea of making available the \proglang{C} code programmed for the work \cite{Ga-DoMa-Be13} to solve exactly moderate to large variable selection problems. Since then, six versions have followed with new abilities that make up the complete toolbox that we have described in this paper.

Nevertheless, \ourpack \ is an ongoing project that we plan as solid and contrasted methods become available. The emphasis is placed on the prior distribution that should be used, since this is a particularly relevant aspect of model selection/testing problems. 

New functionalities that we expect to incorporate in the future are:
\begin{itemize}
\item The case where $n<p+p_0$ and possibly $n<<p+p_0$,
\item specific methods for handling factors,
\item heteroscedastic errors,
\item other types of error distributions.
\end{itemize}

\section*{Acknowledgments}  This paper was supported in part by the Spanish Ministry of Economy and Competitivity under grant MTM2016-77501-P (BAiCORRE). The authors would like to thank Jim Berger for his suggestions during the process of building the package and would like to thank Dimitris Fouskakis for a careful reading and suggestions on this manuscript.
\bibliography{$HOME/Dropbox/BMAinR/bibliography/BMAinR_complete}
\bibliographystyle{plainnat}

\section*{Appendix: Model selection priors for parameters within models}
A key technical component of Bayes factors and hence of posterior probabilities is the prior distribution for the parameters within each model. That is, the prior $\pi_\gamma(\n\alpha,\n\beta_\gamma,\sigma)$ for the specific parameters of the model
\begin{equation}\label{eq.gamma}
M_\gamma:\n y=\n X_0\n\alpha+\n X_\gamma\n\beta_\gamma+\n\varepsilon,\,\,\,\n\varepsilon\sim N_n(\n 0,\sigma^2\n I_n).
\end{equation}
In \ourpack \ the prior used is specified in main functions \code{Btest}, \code{Bvs}, \code{PBvs} and \code{GibbsBvs} with the argument \code{prior.betas} with default value ''Robust'' that corresponds to the proposal the same name in (\cite{Baetal11}). In this paper it is argued, based on foundational arguments, that the robust prior is an optimal choice for testing in linear models.

The robust prior for $M_\gamma$ can be specified hierarchically as
\begin{eqnarray}\label{eq.ourprior}
\pi_\gamma^R(\n\alpha,\n\beta_\gamma,\sigma)=\sigma^{-1}\,N_{p_\gamma}(\n\beta_\gamma\mid \n 0,g\,\n\Sigma_\gamma)
\end{eqnarray}
where $\n\Sigma_\gamma=\sigma^2\, (\n V_\gamma^t\n V_\gamma)^{-1}$, with
\begin{equation}
\label{eq:Vmatrix}
\n V_{\gamma} = (\n I_{n}-\n X_{0}(\n X_{0}^{t}\n X_{0})^{-1}\n X_{0}^{t})\n X_{\gamma}
\end{equation}
and
\begin{equation}\label{eq:piR-g}
g\sim p_\gamma^{R}(g)=\frac{1}{2}\sqrt{\frac{1+n}{p_\gamma+p_0}}(g+1)^{-3/2},\,\,\,\,\,\,g>\frac{1+n}{p_\gamma+p_0}-1.
\end{equation}
For the null model the prior assumed is $\pi_0(\n\alpha,\sigma)=\sigma^{-1}$.

The idea of using the matrix  $\n\Sigma_\gamma$ to scale variable selection priors dates back to \cite{ZellSiow80} and is present in other very popular proposals in the literature. As we next describe, these proposals differ about which distribution should be used for the hyperparameter $g$. Many of these can be implemented in \ourpack \ through the argument \code{prior.betas}.

\begin{itemize}
\item \code{prior.betas="ZellnerSiow"} (\cite{Jef61,ZellSiow80,ZellSiow84}) corresponds to $g\sim IGa(1/2,n/2)$ (leading to the very famous proposal of using a Cauchy).

\item \code{prior.betas="gZellner"} (\cite{Zellner86, KassWass95}) corresponds to fixing $g=n$ (leading to the so called Unit Information Prior).

\item \code{prior.betas="FLS"} (\cite{FLS01}) corresponds to fixing $g=\max\{n,p^2\}$.

\item \code{prior.betas="Liangetal"}  (\cite{liang08}) corresponds to $g\sim\pi(g)\propto (1+g/n)^{-3/2}$.
\end{itemize}

\end{document}